\documentclass[epjCONF,columns]{svjour}

\usepackage{graphics}
\usepackage[varg]{txfonts} 
\usepackage[latin1]{inputenc}

\usepackage{longtable}
\usepackage{url}
\usepackage{rotating}
\usepackage{pslatex}
\usepackage{a4wide}
\usepackage{setspace}
\usepackage{subfigure}
\usepackage{color}
\usepackage{multirow}
\usepackage{fancyhdr}
\usepackage[small]{caption}  
\usepackage{eso-pic}
\usepackage{graphicx}
\usepackage{type1cm}
\usepackage{lineno}

\session-title{Hadron Collider Physics symposium (HCP 2011)}

\title{A Search for Heavy Resonances in the Dilepton Channel}

\begin{document}
\unitlength = 1mm

\author{Daniel Hayden\thanks{\email{daniel.hayden@cern.ch}} On behalf of the ATLAS Collaboration}

\institute{Royal Holloway, University of London.}

\abstract{
There are many extensions to the Standard Model of particle physics which predict the addition of a U(1) symmetry, and/or extra spatial dimensions, which give rise to new high mass resonances such as the Z$'$ and Randall-Sundrum graviton.  The LHC provides a unique opportunity to explore the TeV scale where these phenomena may become apparent, and can be searched for using the precision tracking and high energy resolution calorimetry of the ATLAS detector.  This poster presents the search for high mass resonances in the dilepton channel, and was conducted with an integrated luminosity of 1.08/1.21~fb$^{-1}$ in the dielectron/dimuon channel respectively, at a centre of mass energy $\sqrt{s}$ = 7~TeV.
} 

\maketitle

\section{Introduction}

There are many possible extensions to the Standard Model (SM) predicted at the TeV energy scale which may be visible at the LHC. Many of these extensions predict extra U(1) symmetry with an associated spin-1 particle~\cite{Theory:ZP1}~\cite{Theory:ZP2}. In its simplest form this U(1) symmetry can be arbitrarily added to the existing SM gauge group, resulting in SU(3) $\times$ SU(2) $\times$ U(1) from the SM, and additionally U(1)$^{\prime}$ for the Sequential Standard Model Z$^{\prime}_{SSM}$. More rigorously motivated models proceed via the decomposition of Grand Unified Theories such as E$_6$ $\rightarrow$ SO(10)$\times$U(1)$_{\psi}$ $\rightarrow$ SU(5) $\times$ U(1)$_{\chi}$ $\times$ U(1)$_{\psi}$ leading to Z$^{\prime}$($\theta$) = Z$^{\prime}_{\chi}$cos$\theta$ + Z$^{\prime}_{\psi}$sin$\theta$, where the mixing angle $\theta$ determines the coupling to fermions and results in various possible models with specific Z$^{\prime}$ states. Other extensions of the SM seek to answer questions such as the hierarchy problem where the relative weakness of gravity compared to the other forces of nature can be explained with the use of warped extra dimensions in theories such as the Randall-Sundrum model~\cite{Theory:G}.  A feature of this theory would be a massive spin-2 particle called the graviton (G$^*$) which should be observable at the LHC and have a mass/width that depends on the curvature of the warped dimension, k, and the reduced Planck scale, $\overline{M}_{Pl}$, leading to another parameter of interest, the coupling k/$\overline{M}_{Pl}$.  Both of the new particles mentioned would appear as resonances in the dilepton invariant mass spectrum measured by the ATLAS detector~\cite{ATLAS}, and these results comprise a search using the detector in this endeavor.

\section{Dilepton Resonance Search}

The search for dilepton resonances was conducted in both the electron and muon channels separately, which were then combined to give the final result.  To identify candidate events from data, each analysis selected high energy electron/muon pairs.  The main background to a Z$^{\prime}$/G$^*$ search in these channels is from Drell-Yan, with smaller contributions from $t\bar{t}$, W+jets, diboson, and QCD events~\footnote{QCD events here are defined as semi-leptonic decays of b and c quarks in the dimuon sample, or at least one electron coming from photon conversions, semi-leptonic heavy quark decays or a hadronic fake, in the dielectron sample.}.  These SM background contributions were estimated using Monte Carlo (MC) simulation, except for QCD which was estimated from data using a reverse identification selection sample for electrons, and a non-isolated sample for muons.

For both the dielectron and dimuon channel analyses, a data quality requirement is made to ensure parts of the ATLAS detector important for e/$\gamma$ or $\mu$ analysis respectively are working optimally.  The events are also required to have at least one primary vertex with greater than two tracks, and pass a single electron trigger with a transverse energy (E$_T$) greater than 20~GeV or equivalently for the dimuon analysis, a muon trigger with transverse momentum (p$_T$) greater than 22~GeV.

For an event to be accepted by the analysis in the dielectron channel, an event must contain at least two electron candidates with E$_T$ $>$ 25~GeV and $|$$\eta$$|$ $<$ 2.47, also excluding the region between the barrel and endcap calorimeters 1.37 $\le$ $|$$\eta$$|$ $\le$ 1.52.  The electron candidates that pass these criteria must have been reconstructed from electromagnetic cells clusters with an associated charged particle track from the inner detector. Shower shape variables and hadronic calorimeter leakage, along with information from the inner detector is then used to strengthen the identification of the electron candidates. A hit in the first layer of the pixel detector is required to suppress background from photon conversions.  From the remaining electron candidates the highest E$_T$ pair is selected and the higher E$_T$ electron required to pass an isolation threshold of less than 7~GeV in a cone of 0.2 around the cluster ($\Delta$R = $\sqrt{(\Delta\eta)^2 + (\Delta\phi)^2}$) to reduce the QCD background.  Finally, the invariant mass of the selected pair must be greater than 70~GeV to be accepted by the analysis, and no opposite charge requirement is made to minimise the impact of possible charge mis-identification.

In the dimuon channel, two oppositely charged muons are required.  Each muon must have p$_T$ $>$ 25~GeV, and pass quality criteria from the inner detector as well as having at least three hits in each of the inner, middle, and outer layers of the muon spectrometer to improve momentum resolution. Muons are discarded if they have hits in both the barrel and endcap regions because of residual misalignment. To suppress the cosmic ray background the $z$ position of the primary vertex is required to be less than 200~mm, and muon tracks must have a transverse impact parameter $|$$d_0$$|$ $<$ 0.2~mm, also being within 1~mm of the primary vertex along the beam-line. To reduce the QCD background in the muon channel, each muon is required to be isolated such that $\sum{p_T}$($\Delta$R $<$ 0.3)/p$_T$ $<$ 0.05.  The two highest $p_T$ muons passing this selection form a pair and are required to have an invariant mass greater than 70~GeV to be accepted by the dimuon analysis. 

\begin{figure}
\centering
\subfigure{
\includegraphics[width=75mm]{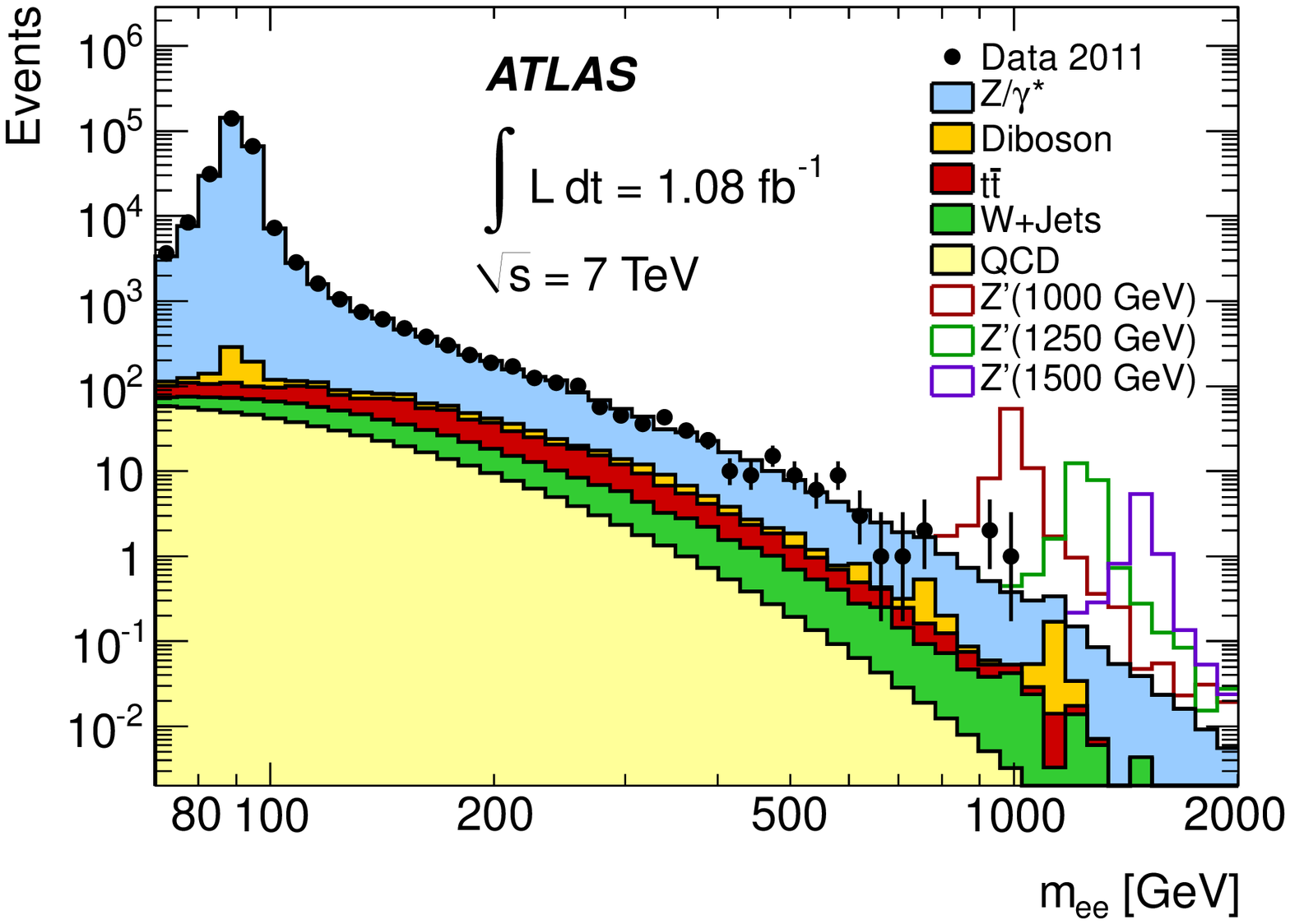}
}
\subfigure{
\includegraphics[width=75mm]{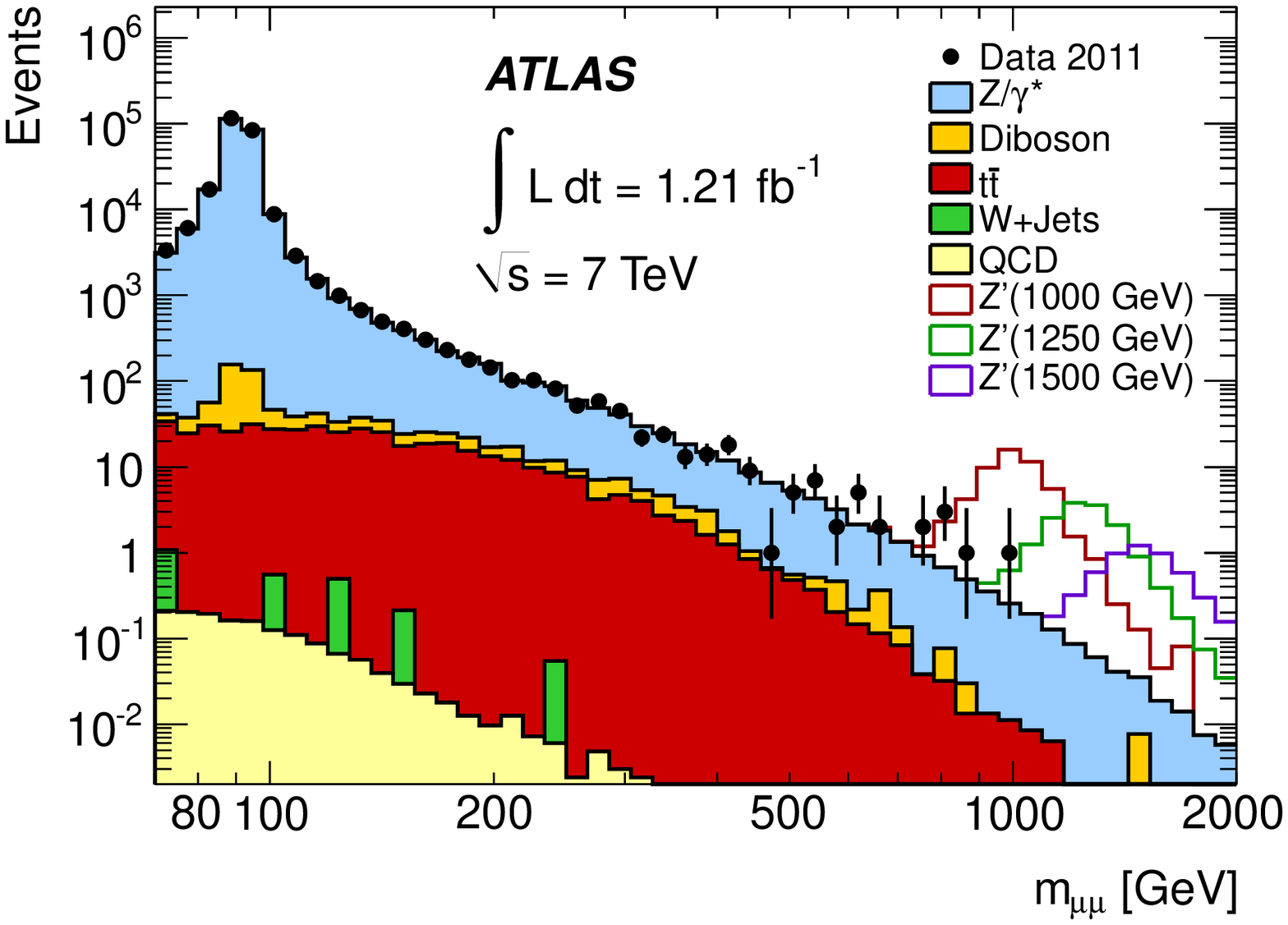}
}
\caption{Invariant mass spectrum for the electron (top) and muon (bottom) channel dilepton resonance search. Various possible Z$^{\prime}_{SSM}$ signals are overlayed to show how an expected signal would manifest itself.}\label{fig:invmass}
\end{figure}

The dilepton analysis was performed with an integrated luminosity of 1.08~fb$^{-1}$ in the electron channel, and 1.21~fb$^{-1}$ in the muon channel.  The results of this analysis can be found in~\cite{Exotics:EPS}, and the main kinematic plots of interest, namely the invariant mass spectrum for both the electron and muon channel, are presented in Figure~\ref{fig:invmass}.

\section{Statistical Analysis}

 Any excess in the observed data over the SM prediction can be quantified using a Log Likelihood Ratio (LLR) test:

\begin{equation}
LLR = -2ln\frac{{\cal L}(data|N_{sig}+N_{bkg})}{{\cal L}(data|N_{bkg})}
\label{eq:LLR}
\end{equation}

In this dataset the greatest excesses give $p$-values of 54\% and 24\% for the dielectron and dimuon channels respectively.  Therefore as no significant excess is observed, limits are set on the cross section times branching ratio ($\sigma$B) for the Z$^{\prime}_{SSM}$ and G$^*$ decaying to leptons, at 95\% confidence level using the Bayesian Analysis Toolkit (BAT)~\cite{BAT}.  BAT constructs a binned likelihood, combining the electron and muon channel searches and accounting for observed ($n$) and expected ($\mu$) events with associated nuisance parameters ($\theta$) on a bin by bin basis:

\begin{equation}
{\cal L}(data|\sigma B,\theta_i) = \prod_{l=1}^{N_{channel}} \prod_{k=1}^{N_{bin}} \frac{\mu_{lk}^{n_{lk}} e^{-\mu_{lk}}}{n_{lk!}}\prod_{i=1}^{N_{sys}}G(\theta_i,0,1)
\label{eq:Likelihood}
\end{equation}

Employing Bayesian statistics (assuming a flat positive prior so that $\pi$($\sigma$B) = 1) and treating the nuisance parameters as Gaussian priors, Markov Chain Monte Carlo is used to reduce the likelihood (${\cal L}^{\prime}$) and obtain the marginalised posterior probability, which is then solved for ($\sigma$B)$_{95}$:

\begin{equation}
0.95 = \frac{\int_0^{{\sigma B}_{95}} {\cal L}^{\prime}(\sigma B)\pi(\sigma B)d(\sigma B)}{\int_0^{\infty} {\cal L}^{\prime}(\sigma B)\pi(\sigma B)d(\sigma B)}
\label{eq:solve}
\end{equation}

\begin{figure}
\centering
\subfigure{
\includegraphics[width=75mm]{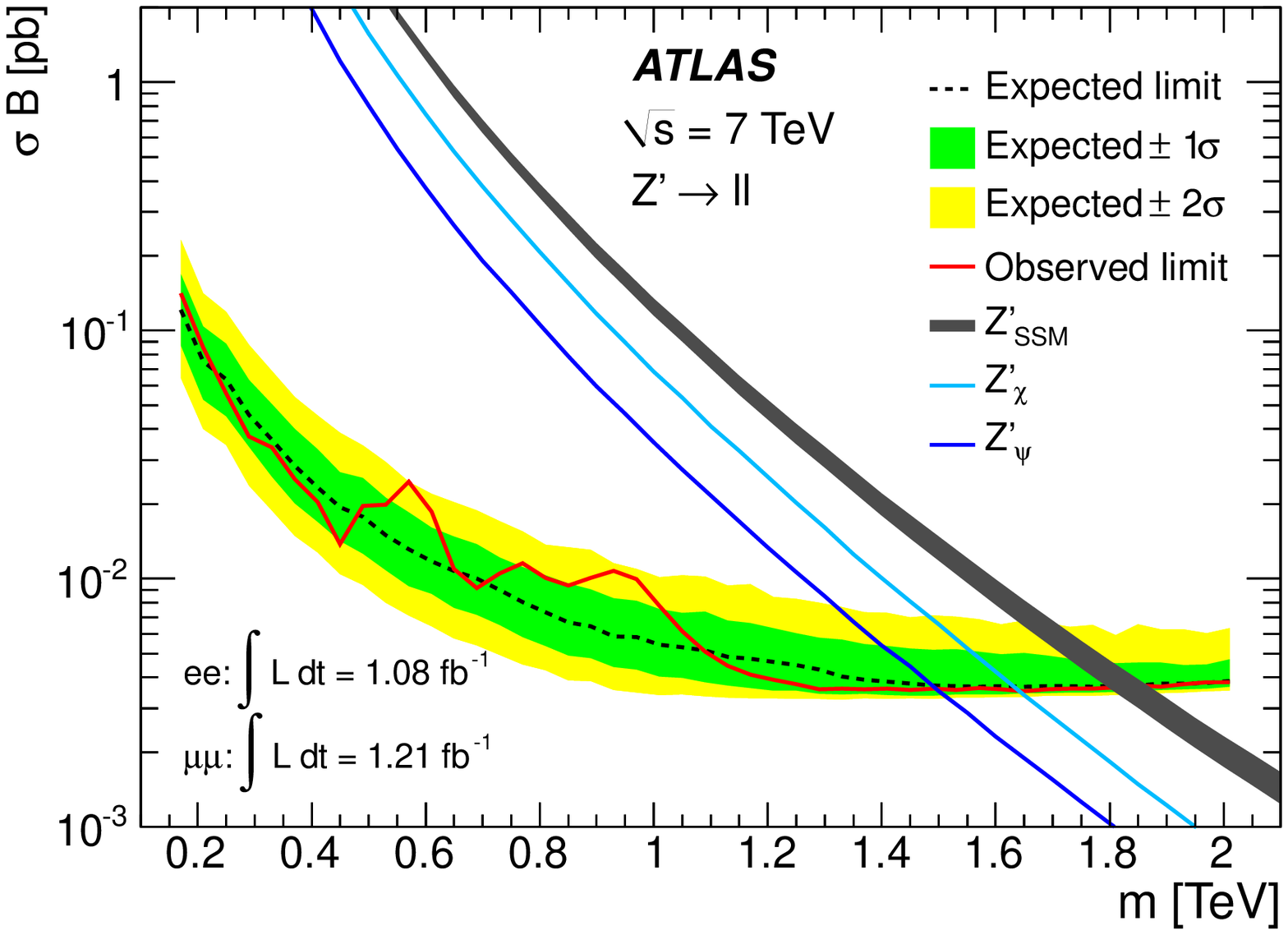}
}
\subfigure{
\includegraphics[width=75mm]{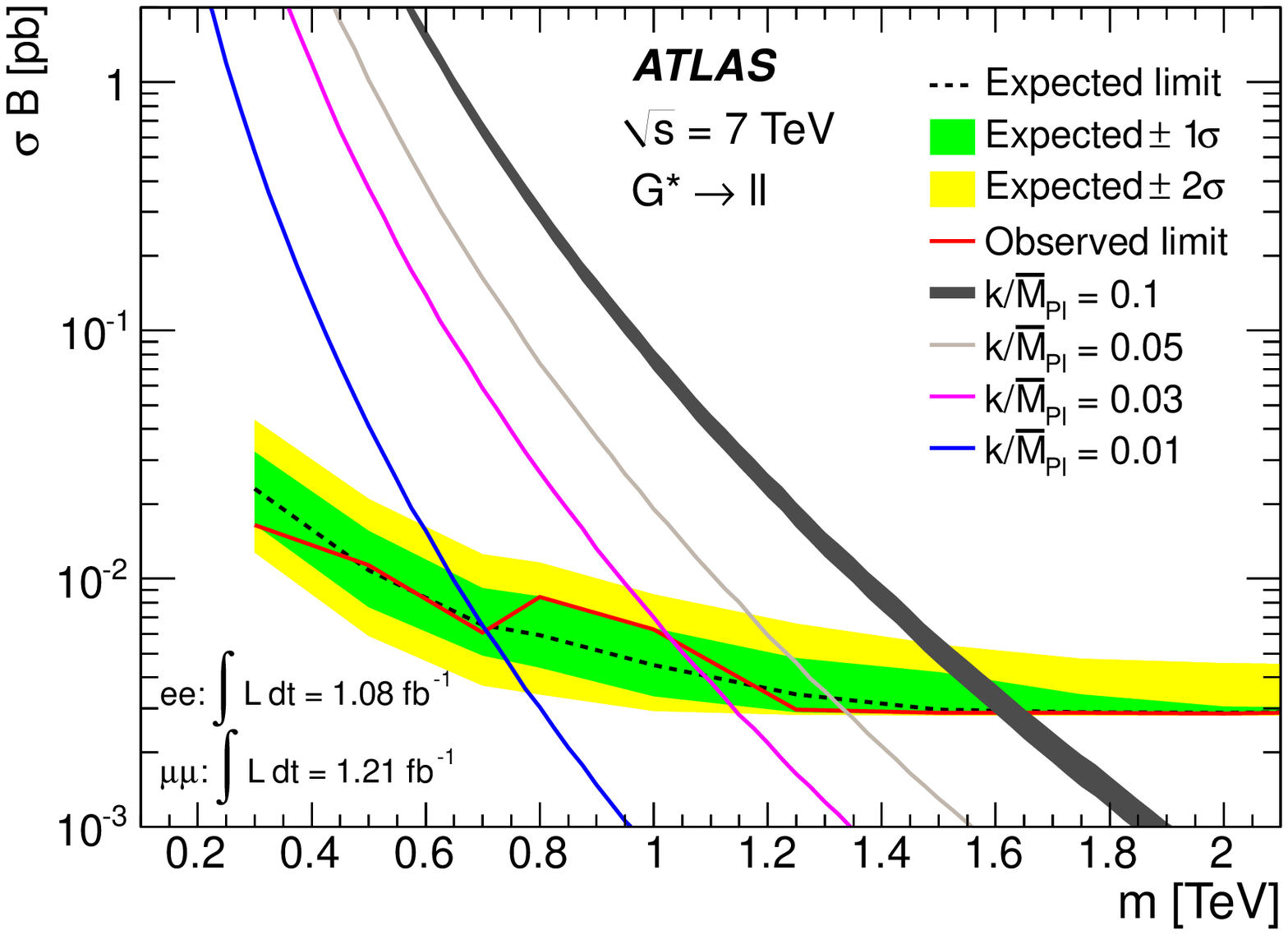}
}
\caption{95\% confidence level $\sigma$B limits for various Z$^{\prime}$ models (top), and RS graviton k/M$_{pl}$ couplings (bottom).}\label{fig:Limits}
\end{figure}

The resulting limits on the Z$'$/G$^*$ $\sigma$B are converted into mass exclusion limits using the theoretical dependence of $\sigma$B as a function of resonance mass.  The $\sigma$B limits are presented in Figure~\ref{fig:Limits}. Table~\ref{tab:Limits} summarises the excluded mass values for the models considered.  The results presented here represent a large step forward in the search for heavy dilepton resonances, exceeding previous experiments' mass exclusion limits for Z$^{\prime}$/G$^*$ resonances in the dilepton channel. With a total integrated luminosity of $\sim$5~fb$^{-1}$ recorded by the ATLAS detector in 2011, this search will soon be updated probing even further into the TeV scale regime in search of new physics beyond the current SM.

\begin{table}[!hbt]
\begin{center}
\begin{tabular}{|c|cccc|}
\hline
\hline
	& \multicolumn{4}{c|}{E$_6$ Z$^{\prime}$ Models} \\
\hline
Model            & Z$^{\prime}_{\psi}$ & Z$^{\prime}_{N}$  & Z$^{\prime}_{\eta}$ & Z$^{\prime}_{I}$  \\
\hline                                                           
Mass limit [TeV] & 1.49 & 1.52 & 1.54 & 1.56 \\
\hline
\end{tabular}

\vspace{0.1cm}

\begin{tabular}{|c|cc|c|}
\hline
	& \multicolumn{2}{c|}{E$_6$ Z$^{\prime}$ Models} & \\
\hline
Model            & Z$^{\prime}_{S}$ & Z$^{\prime}_{\chi}$ & Z$^{\prime}_{SSM}$ \\
\hline                                                           
Mass limit [TeV] & 1.60 & 1.64 & 1.83 \\
\hline
\end{tabular}

\vspace{0.1cm}

\begin{tabular}{|c|cccc|}
\hline
G$^*$ Coupling k/M$_{Pl}$     & 0.01 & 0.03  & 0.05 & 0.10 \\
\hline                                                           
Mass limit [TeV] & 0.71 & 1.03 & 1.33 & 1.63 \\
\hline
\hline
\end{tabular}
\caption{95\% confidence level lower mass exclusion limits for various Z$^{\prime}$ models and RS graviton k/M$_{pl}$ couplings, decaying to two leptons (dielectron or dimuon).}
\label{tab:Limits}
\end{center}
\end{table}

\bibliographystyle{BibStyles/elsarticle-num}
\bibliography{myBib}

\end{document}